\documentclass[12pt,preprint]{aastex}
\received{}
\accepted{}
\slugcomment{Astronomical Journal, 2006Dec26}
\shorttitle{Search for Radio SNe in WR Galaxies}
\shortauthors{Ulvestad et al.}
\journalid{}{}
\articleid{}{}
\def\H2{\ion{H}{2}}

\begin{document}

\title{A VLBI Search for Radio Supernovae in Wolf Rayet Galaxies}

\author{James S.~Ulvestad}
\affil{National Radio Astronomy Observatory} 
\affil{P.O. Box O, Socorro, NM 87801}
\email{julvesta@nrao.edu}
\author{Kelsey E.~Johnson}
\affil{Department of Astronomy, University of Virginia}
\affil{P.O. Box 3818, Charlottesville, VA, 22903}
\email{kej7a@virginia.edu}
\and
\author{Susan G.~Neff}
\affil{NASA's Goddard Space Flight Center}
\affil{Code 681, Greenbelt, MD 20775}
\email{Susan.Neff@gsfc.nasa.gov}

\begin{abstract}

We have used the VLBI High Sensitivity Array (VLBA with VLA,
Green Bank Telescope, and Arecibo) to search for young radio
supernovae in three nearby Wolf Rayet dwarf galaxies containing Super
Star Clusters (SSCs) and signs of extreme star formation in
the last few million years. No milliarcsecond radio sources
were detected in II Zw 40, He 2-10, or NGC 5253, implying
that these galaxies contain few radio supernovae, despite
the fact that they have at least some star formation going
back to 10 million years ago. The upper limits of the source
powers range from $\sim 6\times 10^{17}$ W~Hz$^{-1}$ to $\sim
2\times 10^{18}$ W~Hz$^{-1}$ at 5 GHz, roughly 0.6 to 2.2 times the
power of the galactic supernova remnant Cassiopeia A.
Comparison with the radio supernova population in Arp 299
implies that the current supernova rates in the three dwarf
galaxies are below $10^{-2}$~yr$^{-1}$, consistent with standard
star formation models that predict supernova rates of
$10^{-3}$~yr$^{-1}$ or less in our targets. In He 2-10, the 
VLBI non-detection of a compact VLA source with a significant
nonthermal component indicates that this source may be dominated
by one or more supernova remnants with ages of a few hundred
years or more, which are comparable to Cas~A in power,
and size.

\end{abstract}


\section{Introduction}
\label{sec:intro}

Massive star formation in the Milky Way and in external galaxies
takes place in a wide range of circumstances.  In the current
Universe, perhaps the most extreme star formation takes place
in ``Super Star Clusters'' (SSCs), which may be the forerunners
of globular clusters (e.g., Whitmore 2002).  Individual
SSCs may have ionizing fluxes equal to that of 
$10^3$ to $10^4$ O7-equivalent stars within diameters of just
a few parsecs \citep{whi99,tur00,joh03,tur04}.  The initial 
mass functions in SSCs have been reported to be top-heavy;
however, studies of initial mass functions are affected by
crowding in the fields and mass segregation in clusters \citep{whi02,elm05}.  
Therefore, the total masses of SSCs may be uncertain by an order 
of magnitude, but it appears that at least some SSCs have 
total masses of $10^5M_\odot$ to $10^6M_\odot$.  If they are in 
Wolf-Rayet galaxies, such as the galaxies that are the subjects 
of this paper, the SSCs may form most of their mass into stars 
on a time scale of 1--5$\times 10^6$~yr \citep{kru92}.
Thus their star formation rates per unit area may be above 
$10^{-2}M_\odot$~yr$^{-1}$~pc$^{-2}$, or 
$10^{4}M_\odot$~yr$^{-1}$~kpc$^{-2}$, a factor of at least 1000
higher than the average value in larger-area starbursts
\citep{hec05}.

There are a variety of means of estimating the stellar content, 
star formation rate, or supernova rate in starbursts.  For example, 
one may count ionizing photons by measuring the H$\alpha$ luminosity 
in a starburst galaxy or an SSC.  However, such a method may
greatly underestimate the number of ionizing photons, if the SSC
is heavily obscured by dust.  A more reliable method of counting
ionizing photons may be from the study of thermal radio emission,
although this also may be subject to free-free absorption, leakage
through a porous interstellar medium, or absorption of ionizing
photons by dust.  Mid-infrared imaging also may measure the total
energy output in an SSC, although the resolution of such imaging
often is not adequate to isolate individual SSCs.  Globally, either
far-infrared or radio powers of galaxies may suffice to estimate
their total star-formation rate \citep{con92}, but these data generally are
of resolution too low to isolate individual SSCs.  Yet another
method is to search for young supernovae, thus inferring a supernova
rate that may give clues about the star-formation rate within an
SSC.  However, the supernovae in the densest star-forming regions
may be heavily obscured by dust and impossible to detect optically.
In this circumstance, young radio supernovae may be detected more
easily at centimeter wavelengths; milliarcsecond resolution may be used
to separate them from the more diffuse thermal and
synchrotron radiation present in the SSCs \citep{nef04,rov05,lon06}.

We have made VLBI High Sensitivity Array (HSA) observations
of three nearby dwarf galaxies, II~Zw~40, He~2-10, and NGC~5253, 
containing either one or several
SSCs that are separated by a few parsecs to tens of parsecs.  Existing
mid-infrared and radio interferometric images of these galaxies are
sufficient to isolate the individual SSCs and estimate their total
ionizing fluxes.  All three target galaxies display
significant Wolf-Rayet spectral features, indicating the presence of
(at least) hundreds of massive stars that have evolved off the main
sequence.  Our HSA imaging was carried out in an effort to detect
young radio supernovae within the SSCs in the dwarf galaxies, and
hence to assess the local supernova rates.  This paper reports the
results of our HSA supernova search, which reaches to detection
thresholds near the power of Cas~A in all three target galaxies,
and its implications for their stellar contents and supernova rates.

\section{Observations and Data Reduction}
\label{sec:data}

We observed our three target galaxies on 2005 FEB 26/27 using the HSA
(program ID BU030).  The HSA observations made use of
the 10 antennas of the Very Long Baseline Array (VLBA), 25 phased antennas 
of the Very Large Array (VLA), and the 100m Green Bank Telescope (GBT).  
The 305m Arecibo telescope also was used for II Zw 40, but He 2-10 and 
NGC 5253 are too far south to be accessible to Arecibo.  Each target
was observed in dual-circular polarization at 4.991~GHz frequency, with
total bandwidths of 32~MHz at each polarization and 2-bit sampling.
The total observing time for each target was between 2 and 4 hours,
while the on-source integrations ranged from 1 to 2 hours.  Integration 
times were 10\% to 30\% less for GBT, VLA, and Arecibo, since these
telescopes slew more slowly than the VLBA antennas.  All three targets
were known to be quite weak, so they were observed by means of
phase-referencing to an angularly nearby calibrator source, separated
from the target by $0.7^\circ$ to $2.2^\circ$.  Typical cycle times
in minutes were 1-2-1 (including slew times) for the 
calibrator-target-calibrator sequences; an additional 10-20 seconds
was allowed on each calibrator or target for II Zw 40, due to the
slower slew speed of Arecibo.  Details of the observations
for the individual galaxies are given in Table~\ref{tab:obs}.  

All data calibration was carried out in NRAO's Astronomical Image Processing
System, AIPS \citep{gre03}.  Data flagging was done using VLBA monitor data,
estimates of on-source time for the other antennas, and careful inspection
of the data.  Amplitude calibration for the VLBA, GBT, and
Arecibo was done by using the values for the antenna 
gains maintained by telescope personnel as well as measurements of
system temperature made frequently during the observations; this calibration is 
believed to have an accuracy of 5\%.  The VLA antennas were made mutually
coherent by adjusting their individual phases during the observations of
the phase-reference sources, and the VLA amplitude response was calibrated with
respect to J0137+3429 (3C~48) and J1331+3030 (3C~286), 
using the scale of \citet{baa77}.

For each calibrator and galaxy, a-priori ionospheric corrections were
applied by means of interpolation among models tabulated for every two
hours at regular spacings around the sky, and the most 
accurate available Earth Orientation Parameters were applied,
correcting the less accurate values used at the time of correlation.
Clock and related offsets for the entire observing session were calibrated 
by using observations of the calibrators J0552+0313, J0854+2006, and 
J0927+3902.  Corrections to time-dependent delay and phase errors (largely
due to the atmosphere above each station) were computed using the local
phase-reference sources, and applied to the target galaxies.  For II Zw 40
and NGC 5253, the phase-reference sources were dominated by point sources,
so point-source models sufficed for convergence of this process.  For
He 2-10, the reference source J0846$-$2610 was found to be a well-resolved
double source with a separation of about 40 milliarcseconds, so a 
self-calibrated image of this source was used as a model for the time-dependent
corrections.  Because J0846$-$2610 was completely resolved away on all
baselines to the Mauna Kea VLBA antenna, all Mauna Kea data were flagged
bad for the galaxy He 2-10.  Finally, bad weather above the Hancock VLBA 
antenna led to a 50\% correction to its amplitude calibration on 
II Zw 40, and the antenna was flagged bad on He 2-10.  Because of the
presence of the large GBT and VLA telescopes, the absence of Mauna Kea
and Hancock for II Zw 40 led to little degradation in the achieved 
sensitivity.

\section{Radio Imaging}
\label{sec:imaging}

\begin{figure}
\vspace{18cm}
\includegraphics{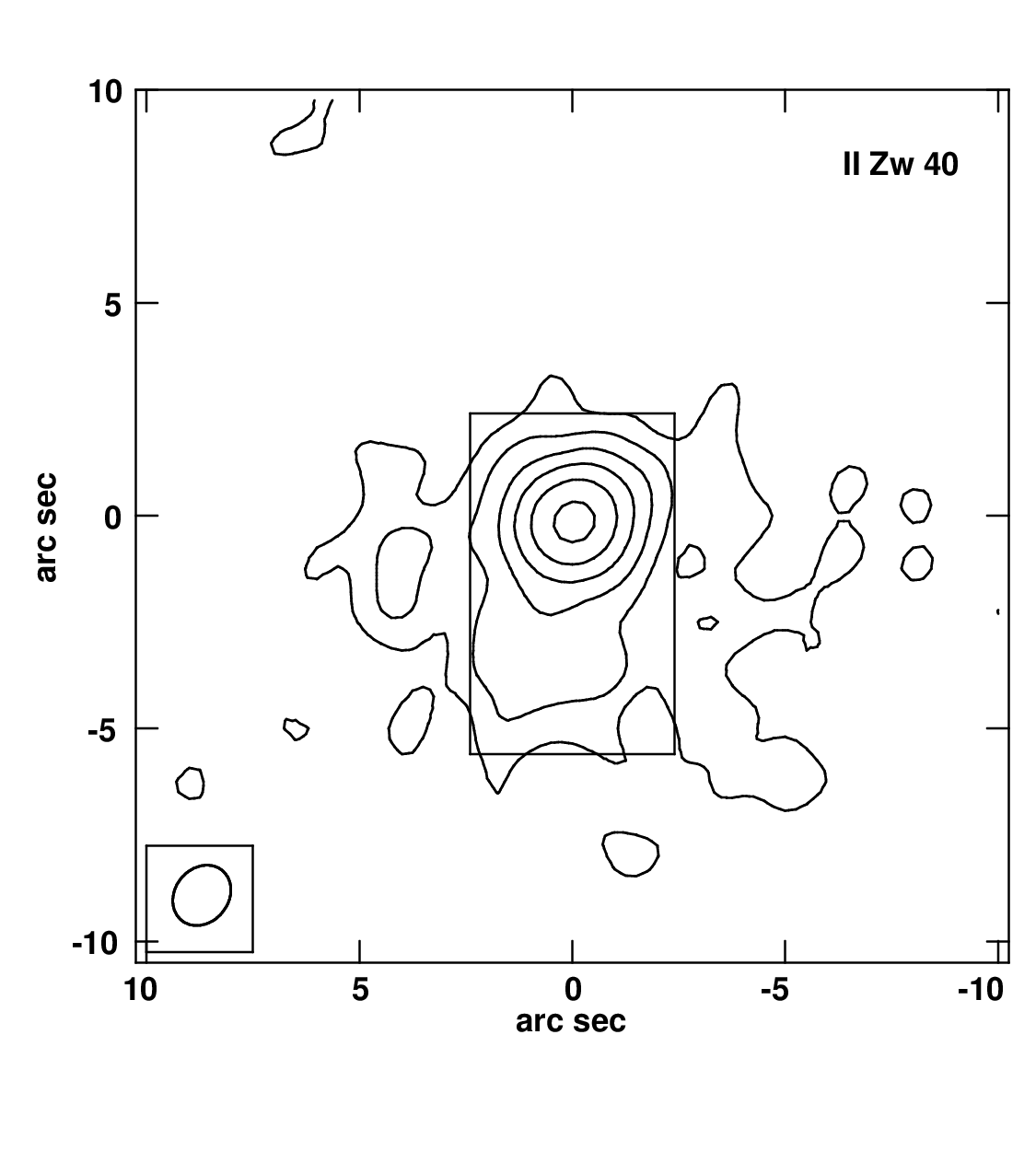}
\includegraphics{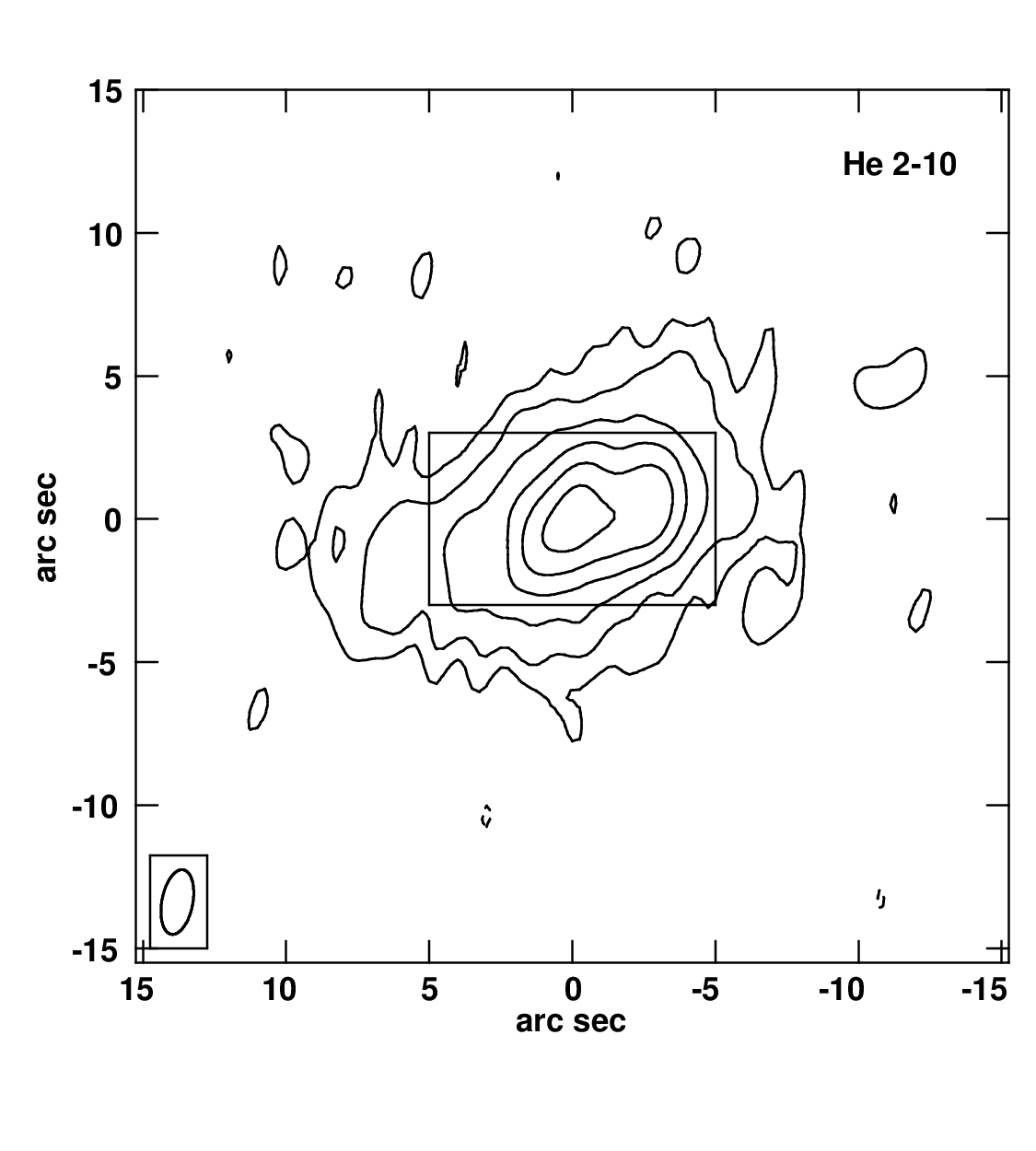}
\includegraphics{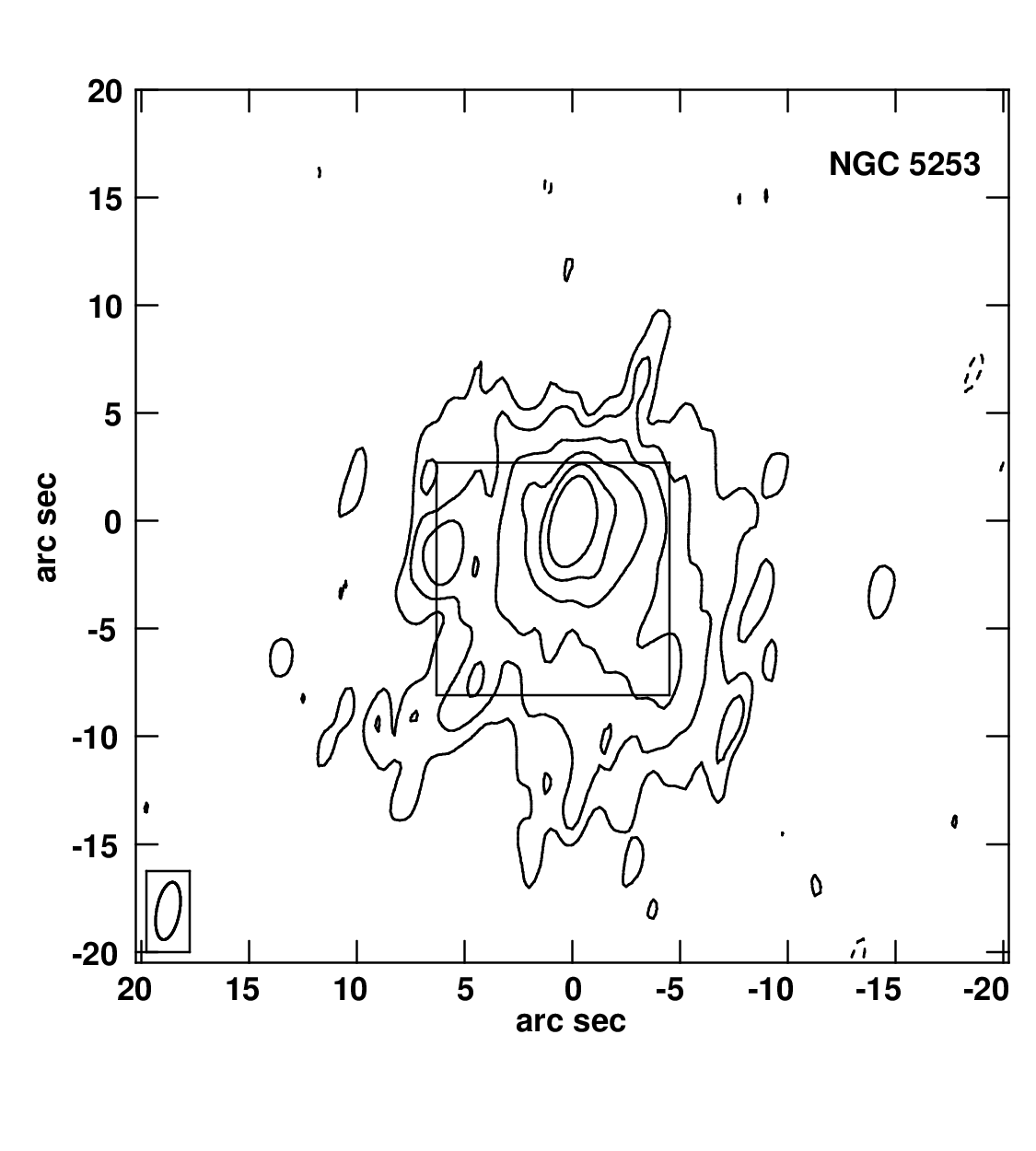}
\caption{
VLA {\bf B} configuration images of three target galaxies at 4.99 GHz, made 
from the phased VLA observations obtained on 2005 FEB 26/27.  From top to bottom,
the images are of the galaxies II Zw 40, He 2-10, and NGC 5253, respectively.
The lowest contour in each case is 3 times the rms noise, with successive
contours increasing by factors of two (negative contours are shown dashed).
The synthesized beam is shown in the lower left corner of each image, and
the box shown near the center of each image indicates the region imaged with
the HSA to search for milliarcsecond radio sources.  More detailed
parameters of the images are given in Table~\ref{tab:vla}. }
\label{fig:vla}
\end{figure}

As part of the imaging process, we made VLA images of all three
target galaxies, which enabled us to confirm the locations where we
might search for compact milliarcsecond-scale radio emission in the HSA
data.  Since the VLA was in its {\bf B} configuration at the time of
our observations, with a maximum baseline length of 9~km, the resolution 
of these VLA observations was roughly 1.5~arcseconds.  Figure~\ref{fig:vla}
shows the resulting VLA images at 4.99 GHz; a few details of the images are
listed in Table~\ref{tab:vla}.

The VLBA images were made with ``natural'' weighting of the visibility
data in order to achieve the best possible sensitivity.  
The imaging was carried out over a number of 4096 by 4096 pixel
fields, with pixel sizes of 0.4--0.5 milliarcseconds.  The 
resolution of the images was sub-parsec (see Table~\ref{tab:obs}), 
typically with better resolution in the east-west direction
since the sources were at low declination and
the VLBI array had a significant east-west elongation.  Spectral and 
time averaging in the correlation and imaging were set so that
the peak flux densities in the relatively wide 
fields imaged were not degraded by averaging of the data (see Bridle \& 
Schwab 1999 for discussion of this effect).  The total area imaged at
milliarcsecond resolution is shown for each galaxy in Figure~\ref{fig:vla}.
Although these relatively low resolution VLA images display some radio emission
outside the areas imaged with HSA, higher resolution images of all galaxies
(published, or made from archival VLA data) 
confirm that all regions with sub-arcsecond 5 GHz emission were
included in the HSA imaging.

For each galaxy, between $10^8$ and $10^9$ pixels were included in the
wide-field HSA images.  Thus, in the absence of a-priori information
about the location of milliarcsecond radio sources, setting a $3\sigma$
source detection threshold clearly is inadequate.  Most individual images 
contained pixels between 5 and 6 times the rms noise, and a few fields had 
pixels above 6 times the noise, in seemingly random locations.  
Therefore, we set conservative upper limits
for each galaxy of 7 times the rms noises given in Table~\ref{tab:obs}.
To get true upper limits, it also is necessary to account for possible
losses of interferometer coherence in the phase-referencing process.
To estimate this coherence loss, we observed the ``check'' source
J0826$-$2230 occasionally in place of He~2-10 and J1316$-$3338 in place
of NGC~5253.  (No check source was observed for II Zw 40 because of
the long slew times that would have been 
required for the Arecibo telescope).  Comparing the
peak flux densities of J0826$-$2230 and J1316$-$3338 in phase-referenced 
images to those derived after self-calibration indicates respective 
coherence loss factors of 2 and 3 at calibrator-galaxy separations of 
5\fdg9 and 3\fdg8.  Coherence losses typically increase approximately
linearly with the calibrator-galaxy separation \citep{bea95}, so we infer
reductions in the peak amplitudes by a factor of 1.4 for He~2-10 and
a factor of 2.0 for NGC~5253.  II~Zw~40 was observed at a much smaller
switching angle of 0\fdg7. and at a higher elevation than the other two
galaxies; averaging the interpolated values for the other two galaxies, 
we estimate a reduction factor of 1.2 for the peak amplitudes of any
real sources in II~Zw~40.  Thus, we arrive at the final upper limits for
each galaxy which are given in Table~\ref{tab:limits}.

\section{Individual Galaxies}
\label{sec:indgal}

\subsection{II Zw 40}
\label{ss:iizw40}

II~Zw~40 is a dwarf starburst galaxy at a distance of 10.5 Mpc \citep{bec02}.
It is known to be a Wolf-Rayet galaxy \citep{kun83,vac92},
suggesting significant star formation 3-6 Myr in the
past and indicating the presence of a population of evolved massive
stars that are temporally associated with the onset of supernovae.
From near-infrared and optical imaging, \citet{van96} 
suggest that the bulk of the recent starburst activity is less than
4~Myr old, which implies that the galaxy still may be ramping up to its
peak number of evolved massive stars and supernovae.  Based on their
models, \citet{van96} predict a supernova rate of
$<10^{-3}$~SN yr$^{-1}$. Neutral hydrogen observations indicate moderate
depressions associated with the active star forming region, plausibly
suggesting the impact of mechanical feedback from stellar winds and/or
supernovae \citep{van98}.  

Previous radio observations of II~Zw~40 indicate a relative lack of
non-thermal emission.  \citet{sra86} find that the 5 GHz
flux is dominated by thermal emission, and estimate a total non-thermal
flux at 1.49 GHz of 2.4--3.9~mJy, which is consistent with II~Zw~40
only having a low level of recent supernovae activity.
\citet{bec02} fit a radio spectrum to II~Zw~40 indicating that most of
the radio emission is optically thin thermal emission, presumably energized
by young stars; from the radio source in the central $3''$, they infer
the presence of $\sim 1.4\times 10^4$ O7-equivalent stars.  Larger scale 
emission inferred by \citet{bec02} is imaged more clearly in our 
Figure~\ref{fig:vla}, but this emission clearly is not compact on 
milliarcsecond scales.  Thus, the region that we imaged using the HSA 
contains all the subarcsecond-scale flux in II~Zw~40.  The most 
compact 15 GHz sources in this region have flux densities of
0.4--0.6~mJy~beam$^{-1}$, implying similar 5~GHz flux densities for thermal
sources, so it would be feasible for these sources
to contain radio supernovae at the level of 0.1--0.3~mJy at 5~GHz.  
Our conservative HSA upper limit of 0.09~mJy~beam$^{-1}$ is somewhat lower,
and corresponds to an upper limit of $1.2\times 10^{18}$~W~Hz$^{-1}$ at 5~GHz.  
We can compare this limit to the 
power of the galactic supernova remnant Cas~A, which presently has a
5~GHz flux density of 650~Jy \citep{baa77}.  For a distance of
3.4~kpc \citep{ree95}, the Cas~A power at 5 GHz is 
$9.0\times 10^{17}$~W~Hz$^{-1}$, so we conclude that there are no radio
supernovae in II~Zw~40 with powers greater than 1.3 times Cas~A.

\subsection{He 2-10}
\label{ss:he210}

He 2-10 is another dwarf galaxy, at a distance of 9~Mpc.  Its 
current star formation is dominated by a star-forming region that
contains several hundred Wolf-Rayet stars \citep{vac92} and 
$H\alpha$ emission energized by $\sim 2\times 10^4$ O7-equivalent stars
\citep{bec99} in a starburst of age 10--20~Myr \citep{bec97}.
A series of high-resolution optical, mid-infrared, and radio imaging
observations \citep{kob99,joh00,vac02} culminated in a high-resolution
VLA image of five compact radio sources spread over about $4''$ \citep{joh03}.
These five radio sources have 5~GHz flux densities ranging from 0.6 to 2.3~mJy;
based on VLA observations ranging from 5 GHz to 43 GHz, \citet{joh03} 
modeled four of them as partially opaque thermal 
bremsstrahlung sources.  Our 5~GHz HSA imaging of the entire region covered 
by the five VLA sources shows no detections of milliarcsecond radio emission, 
with the conservative upper limit of 0.21~mJy~beam$^{-1}$ corresponding to 2.2 
times the power of Cas~A.

VLA source 3 found by \citet{joh03} is of particular interest here because 
its spectrum appears to be dominated by nonthermal emission, as one might 
expect from a young supernova or collection of supernova remnants, or even 
an active galactic nucleus.  Therefore, we have examined the vicinity of this
radio source carefully in our HSA image, searching for an excess of
$5\sigma$ or $6\sigma$ radio sources.   Within 0\farcs25 of source 
3, the peak flux density is only $4.5\sigma$, implying that we
detect no young supernovae within that nonthermal source.
Given rough upper limits of $\sim 0.3$~mJy for the
5~GHz thermal radio emission from this source (estimated by using the 
highest frequency radio emission), its total mass in stars can be
scaled from the thermal models of the other VLA sources \citep{joh03}
and is found to be $\sim 5\times 10^4 M_\odot$.  Hence, the local 
supernova rate within this source is expected to be less than 
$10^{-3}$~yr$^{-1}$.  Our HSA upper limit (taking account of the estimated 
VLBI coherence loss) is 0.13~mJy~beam$^{-1}$, or 1.4 times
the power of Cas~A, while the total VLA nonthermal emission of 
$\sim 0.4$~mJy is about 4 times the Cas~A power.  
The HSA beam size of $12\times 2$~mas 
corresponds to a linear size of roughly $0.5\times 0.1$~pc.
Thus we suggest the possibility that at least half of the 
VLA-scale radio emission in this source is generated by one 
or more supernova remnants with ages of a few hundred to a few
thousand years, total power several times that of Cas~A and 
diameter(s) of 0.5~pc or larger.  Such a source or sources would account 
for the nonthermal emission detected by the VLA, but would be too 
heavily resolved to be detected in our HSA observations.

\subsection{NGC 5253}
\label{ss:n5253}

NGC~5253 also is a Wolf-Rayet galaxy \citep{vac92}, located at a distance
of only 4.1~Mpc \citep{san94}.  The strength of the Br$\alpha$ line indicates
the presence of $10^3$ O7-equivalent stars within the central $2''$, and
$2.5\times 10^3$ such stars in the central $20''$, or 400~pc \citep{cro99}. 
\citet{cal97} used HST imaging to identify several young clusters in the
inner few arcseconds, including two with ages of 2.5 and 3--4~Myr. They find
that the younger cluster may be as massive as $10^6M_\odot$, and the total star
formation intensity averaged over the central $6''$ is 
10--100 $M_\odot$~yr$^{-1}$~kpc$^{-2}$.
Two Type I supernovae brighter than 9th magnitude have been confirmed 
in NGC~5253 in the last 111 years:
SN1895B (Z Cen) occurred $26''$ northeast of the galaxy nucleus
\citep{cam97}, while SN1972E was $102''$ to the southwest \citep{kow72}.
A third reported supernova, SN1986F, apparently was not a real event
\citep{fil86}.  The full region imaged by the VLA in B configuration
during our VLBI observations
(not shown in Figure~\ref{fig:vla}) does not detect either SN1895B or 
SN1972E, with $3\sigma$ upper limits of 0.17~mJy~beam$^{-1}$, considerably
fainter than the 1.4 GHz upper limits for the supernovae that were found by 
\citet{deb73} in the early 1970s.  Both of these supernovae were reported to
be Type I, which typically are undetected (Type Ia) or have rapid turn-on
and turn-off (Type Ib/c) at radio wavelengths, so the lack of any VLA
detection is not surprising.

Diffuse radio emission is seen to span the entire $20''$ central region
\citep{tur98}; both of the confirmed supernovae mentioned above lie
well outside this region.  \citet{tur00} found a ``radio supernebula'' 
in the central arcsecond that is optically thick near 15~GHz, and used the 
radio flux density to infer the presence of $4\times 10^3$ O7-equivalent 
stars in the central few parsecs, considerably higher than suggested from the
Br$\alpha$ line,  At the very center of the starburst, a compact double
radio source was imaged by \citet{tur04} using the VLA at 43 GHz.  This
appears to coincide with a near-infrared double star cluster with
a separation of 6-8~pc \citep {alo04}. The eastern member of that cluster
pair corresponds to the youngest cluster identified by \citet{cal97}, and
also seen in the ultraviolet by \citet{meu95}.

\begin{figure}
\plotone{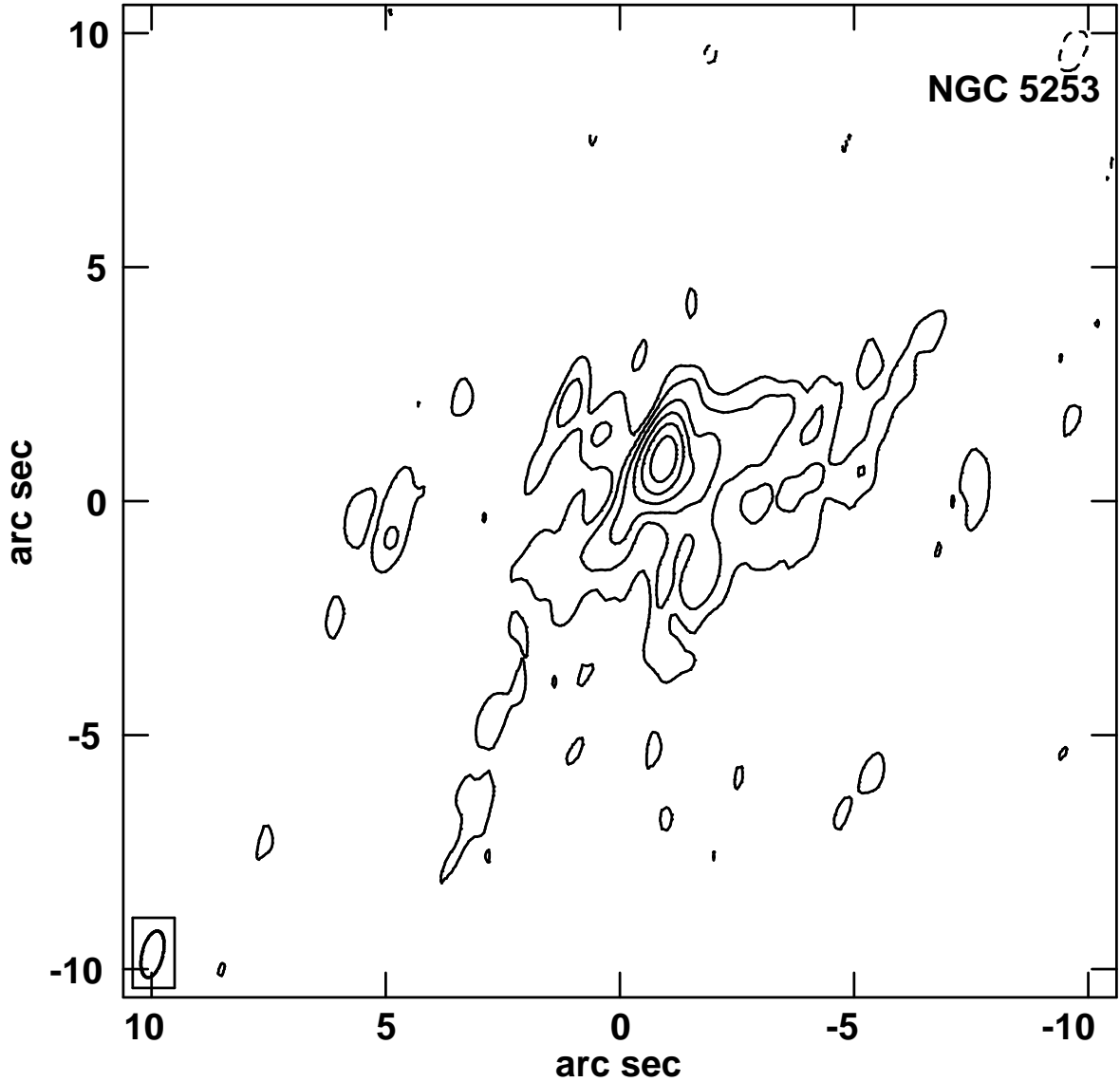}
\caption{
VLA {\bf A} configuration image of NGC 5253 at 5.0 GHz, from archival
data observed on 1986 APR 29, and
imaged with natural weighting of the visibilities.  The lowest contour level 
is at 3 times the rms noise of 52~$\mu$Jy~beam$^{-1}$, with successive 
contours increasing by factors of two (negative contours are shown dashed).
The synthesized beam of $1.02'' \times 0.44''$ in position angle $-14^\circ$
is shown in the lower left.  Coordinates are given as offsets from the J2000 
position of $\alpha$=13h39m56.051s, $\delta=-31^\circ 38'25.22''$. }
\label{fig:a5253}
\end{figure}

Our HSA images cover the $10''$ region of NGC~5253 with the most compact
radio emission, as shown in Figure~\ref{fig:vla}.  Within this region, there
is emission on scales ranging from subarcsecond to many arcseconds.
There are no published 5~GHz images of NGC~5253 made with the VLA
{\bf A} configuration, so we extracted and processed an {\bf A}
configuration data set obtained to search 
for radio emission from the putative supernova 1986F, resulting in the
radio image shown in Figure~\ref{fig:a5253}.  Comparison of this image with 
the {\bf B} configuration image shows that there 
is considerable 5 GHz emission on scales between $0.1''$ and $0.5''$ that is
poorly sampled by our HSA observations; in fact, the effects of this emission
forced us to remove the sensitive baseline between the phased VLA and the
Pie Town VLBA antenna (60~km away) in order to make a noise-limited HSA image.
The inferred 5~GHz upper limit of 0.29~mJy~beam$^{-1}$ on milliarcsecond
scales corresponds to only 0.6 times the radio power of Cas~A.

\section{Starburst Ages and Supernova Rates}
\label{sec:ages}

The limits that we place on the radio supernova populations in our target
galaxies can be converted to radio supernova rates, if a few simplifying
assumptions are made.  Conversion of a radio supernova rate to an overall
supernova rate is most believable
if we can make some estimate of the fraction of supernovae in our galaxies
that should produce detectable radio supernovae.  The radio emission from
young type~II supernovae in ``normal'' galactic environments typically is dominated
by the interaction of the supernova blast wave with the mass-loss shell
from the parent star \citep{che82}.  However, in the highest-density
regions, such as SSCs in dense molecular clouds, the supernova blast may be 
dominated by the high-density interstellar environment \citep{che01}.  In
such a case, the supernova ejecta will be slowed dramatically,
particle acceleration at the shock front will be quite efficient, and
radio emission will be both powerful and long-lived.  Thus, it
appears to be a reasonable inference that {\it every} type~II supernova
in an SSC may produce significant radio emission, 

Predictions of the supernova rates in our target galaxies may be derived 
from their populations of massive stars and the estimated lifetime of a
radio supernova.  The compact starburst regions in our target
galaxies must contain roughly $10^4$ O7-equivalent stars in order 
to generate the ionizing flux that energizes the H$\alpha$ and thermal 
radio emission.  In the nearby galaxies NGC~253 \citep{ulv94,ulv97} and 
M82 \citep{ulv94,kro00}, the radio flux densities of most young supernova 
remnants are relatively constant over periods of 8--15 years, implying 
that they stay near their radio peaks for periods of at least 
20--100 years.  This supports the modeling of \citet{che01}, but is in
contrast to Type II radio supernovae in less dense environments, whose flux
densities fall off with time $t$ as roughly $t^{-0.7}$ to $t^{-1.8}$ 
\citep{wei02}. 

We may use the stellar contents and inferrred radio lifetimes discussed
above to estimate the supernova rates expected in our program galaxies.
Roughly, the main-sequence lifetimes of massive young stars that go 
supernova range from $\sim$3--40~Myr (cf. Starburst99 models of 
Leitherer et al. 1999), and the supernova rate should 
peak about 4--6~Myr after an instantaneous starburst.  Thus, if we 
assume that the massive stars all formed in a time of 1--3~Myr, and that 
there are $\sim 10^4$ such stars, we should expect a radio supernova 
rate no higher than $\sim$1--5$\times 10^{-3}$~yr$^{-1}$
at a time about 5--7$\times 10^6$~yr after the starburst.  Alternatively,
one may take a quasi-instantaneous burst with a total mass of $\sim 10^6 M_\odot$;
the Starburst99 models \citep{lei99} then show a peak supernova rate
of slightly less than $10^{-3}$~yr$^{-1}$ about 6~Myr after the burst.  Thus,
as long as the typical radio supernova stays above $10^{18}$~W~Hz$^{-1}$
for no more than a few hundred years, this is consistent with our lack
of detection of any milliarcsecond radio sources in our three target galaxies.

The upper limits to milliarcsecond-scale radio emission may be compared
directly to galaxies with much higher global star formation rates in 
order to estimate the supernova rates in our galaxies.  The two most
extreme examples are Arp~220 \citep{smi98,rov05,lon06} and Arp~299 
\citep{nef04,ulv05}.  In Arp~299, the primary nucleus (``A'') has
an approximate supernova rate of 0.6~yr$^{-1}$ \citep{alo00}, and
contains $\sim 10$ milliarcsecond radio supernovae with 5~GHz powers in
the range between $10^{19.6}$~W~Hz$^{-1}$ and $10^{20}$~W~Hz$^{-1}$ 
\citep{ulv05}.  If the number of radio supernovae
of flux density $S$ is a power-law $N(S)\propto S^{-\beta}$, then
one immediately finds that the number of radio supernovae in Arp~299-A
with 5~GHz powers above $10^{18}$~W~Hz$^{-1}$ should range from $\sim 24$
for $\beta=0.5$ to $\sim 50$ for $\beta=1.0$.  Our three
target galaxies have no radio supernovae detected above limits ranging 
from $10^{17.7}$ to $10^{18.3}$~W~Hz$^{-1}$.  As a rough estimate
we take an upper limit of 1 supernova above $10^{18}$~W~Hz$^{-1}$
for each galaxy; scaling by the supernova rate in Arp~299, we thus
find upper limits of $\sim 10^{-2}$~yr$^{-1}$ for the supernova rate
in each target galaxy, consistent with the expectation from the
apparent stellar populations.\footnote{It would be possible to make a 
more rigorous statistical derivation of the supernova rate that 
would yield a 95\% likelihood of each galaxy having no supernovae with 
radio powers above its upper limit, but 
this is not warranted given the uncertain coherence correction of the 
observed HSA upper limits, uncertain values of $\beta$, and the strong
dependence of supernova radio powers on the local environments.}

Our results overall are consistent with the hypothesis that the
starburst galaxies containing young radio supernovae or supernova
remnants are those which have star formation over very large extents,
rather than being dominated by one or a few SSCs.  Despite the intense
nature of their star formation in small volumes, dwarf galaxies such as
those observed in this study do not have enough mass to generate
a supernova rate that will lead to detection of young radio supernovae
at present VLBI detection thresholds.

\section{Conclusions}
\label{sec:conc}

We have used the VLBI High Sensitivity Array at milliarcsecond
resolution to search for young radio supernovae in three nearby
dwarf Wolf-Rayet starburst galaxies.  No such supernovae were detected at
radio power levels approximately equal to the power of Cas~A.  The
non-detections are consistent with upper limits of less than 
$10^{-2}$~yr$^{-1}$ for the supernova rates in the dwarf starburst galaxies,
compared to an expected supernova rate of $\sim 10^{-3}$~yr$^{-1}$
for starbursts of $\sim 10^6M_\odot$.  We suggest the possibility
that VLA source~3 in He 2-10 includes one or more supernovae that are 
a few hundred to a few thousand years old, somewhat more powerful than 
Cas~A, and with diameter(s) greater than 0.5~pc, thus
being undetectable in the present HSA observations.  We expect that VLBI 
detection thresholds about 10 times fainter in our program galaxies, 
near 0.1 times the Cas~A power, would begin to show
partially resolved radio emission from supernova remnants with ages
of a few thousand years.

\acknowledgments

The National Radio Astronomy Observatory is a facility of the 
National Science Foundation operated under cooperative agreement by Associated 
Universities, Inc.  We thank the staffs of the VLA, VLBA, GBT, and Arecibo
telescopes that made these observations possible.  KEJ is grateful for support
provided by an NSF CAREER award.  We acknowledge
the authors of Starburst99 for making their modeling software available
on line, as well as the use of the NASA Extragalactic Database (NED).
Thanks to the referee for a careful reading of the manuscript.

{\it Facilities:} \facility{VLBA}, \facility{VLA}, \facility{GBT},
\facility{Arecibo}

\clearpage

\clearpage

\clearpage

\begin{deluxetable}{lccccccc}
\tabletypesize{\scriptsize}
\tablecolumns{8}
\tablewidth{0pc}
\tablecaption{High Sensitivity Array Observations of Wolf Rayet Galaxies}
\tablehead{
\colhead{(1)}& \colhead{(2)}& \colhead{(3)}& \colhead{(4)}&
\colhead{(5)}& \colhead{(6)}& \colhead{(7)}& \colhead{(8)} \\
\colhead{Galaxy}&\colhead{Date}&\colhead{Telescopes}&
\colhead{Integration}&\colhead{resolution}&\colhead{rms noise}&
\colhead{Ref. Source}&\colhead{Separation} \\
\colhead{}&\colhead{}&\colhead{}&\colhead{(min)}&\colhead{(mas$\times$mas)}&
\colhead{($\mu$Jy beam$^{-1}$) }&\colhead{}&\colhead{(deg.)} }
\startdata
II Zw 40&2005 Feb 26/27 & VLBA,VLA,GBT,Arecibo&70&$4.4\times 1.5$&10.5&
J0552+0313&0.7 \\
He 2-10&2005 Feb 27&VLBA,VLA,GBT&124&$12.0\times 1.9$&21.0&J0846$-$2610&2.2 \\
NGC 5253&2005 Feb 27&VLBA,VLA,GBT&126&$9.4\times 1.8$&20.4&J1330$-$3122&2.1 \\
\enddata
\label{tab:obs}
\end{deluxetable}

\clearpage

\begin{deluxetable}{lccccc}
\tablecolumns{6}
\tablewidth{0pc}
\tablecaption{Parameters of VLA {\bf B} Configuration Images of SSC Galaxies\tablenotemark{a}}
\tablehead{
\colhead{(1)}& \colhead{(2)}& \colhead{(3)}& \colhead{(4)}&
\colhead{(5)}& \colhead{(6)} \\
\colhead{Galaxy}&\multicolumn{2}{c}{\underbar{Image Center (J2000)}}&
\colhead{Peak}&\colhead{Resolution}& \colhead{rms Noise} \\
\colhead{}&\colhead{RA}&\colhead{Dec.}& \colhead{}&
\colhead{}&\colhead{} \\
\colhead{}&\colhead{(h m s)}&\colhead{($^\circ\ '\ ''$)}&
\colhead{(mJy beam$^{-1}$)}&\colhead{(arcsec)}&
\colhead{($\mu$Jy beam$^{-1}$)} }
\startdata
II Zw 40&05 55 42.620&\ 03 23 32.10&5.4&$1.51\times 1.23$, pa $-39^\circ$&45 \\
He 2-10&08 36 15.230&$-$26 24 34.20&5.2&$2.29\times 1.08$, pa $-11^\circ$&34 \\
NGC 5253&13 39 55.965&$-31$ 38 24.40&24.4&$2.70\times 1.06$, pa $-10^\circ$&63 \\
\enddata
\tablenotetext{a}{The VLA data were taken from the HSA observations, made on the
dates listed in Table~\ref{tab:obs}.}
\label{tab:vla}
\end{deluxetable}

\clearpage

\begin{deluxetable}{lccccc}
\tabletypesize{\scriptsize}
\tablecolumns{6}
\tablewidth{0pc}
\tablecaption{HSA Upper Limits in SSC Galaxies}
\tablehead{
\colhead{(1)}& \colhead{(2)}& \colhead{(3)}& \colhead{(4)}&
\colhead{(5)}& \colhead{(6)} \\
\colhead{Galaxy}&\colhead{Image rms}&\colhead{$7\sigma$ Limit}&
\colhead{Correction Factor\tablenotemark{a}}&\colhead{Final Limit}&
\colhead{Power Limit} \\
\colhead{}&\colhead{($\mu$Jy beam$^{-1}$)}&
\colhead{($\mu$Jy beam$^{-1}$)}&\colhead{}&
\colhead{($\mu$Jy beam$^{-1}$)}&\colhead{(W Hz$^{-1}$)} }
\startdata
II Zw 40&10.5&73.5&1.2&88&$1.2\times 10^{18}$ \\
He 2-10&21.0&147&1.4&206&$2.0\times 10^{18}$ \\
NGC 5253&20.4&143&2.0&286&$5.5\times 10^{17}$ \\
\enddata
\tablenotetext{a}{The ``Correction Factor'' is derived from an estimate 
of the coherence loss in the phase-referencing process (see text).}
\label{tab:limits}
\end{deluxetable}

\end{document}